\documentclass[%
 reprint,showkeys,
 amsmath,amssymb,
 aps,
]{revtex4-2}

\usepackage{graphicx}
\usepackage{dcolumn}
\usepackage[usenames]{color}
\usepackage{colortbl}
\usepackage{bm}
\usepackage{longtable}

\begin{document}

\title{Isomeric pair ${^{95\rm m,g}\rm{Nb}}$ in photonuclear reactions on $^{\rm nat}$Mo \\ at end-point bremsstrahlung energy of 35--95 MeV} 


\author{I.S. Timchenko$^{1,2,}$} \email{timchenko@kipt.kharkov.ua;\\ iryna.timchenko@savba.sk}
\author{O.S. Deiev$^2$, S.M. Olejnik$^2$, S.M. Potin$^2$, \\
	V.A. Kushnir$^2$, V.V. Mytrochenko$^{2,3}$, S.A. Perezhogin$^2$, V.A. Bocharov$^2$}

\affiliation{$^1$ Institute of Physics, Slovak Academy of Sciences, SK-84511 Bratislava, Slovakia}%
\affiliation{$^2$ National Science Center "Kharkov Institute of Physics and Technology", 1, Akademichna St., 61108, Kharkiv, Ukraine}%
\affiliation{$^3$ CNRS/IJCLAB, 15 Rue Georges Clemenceau Str., Orsay, 91400, France}%

\date{\today}

\begin{abstract}
The ${^{\rm nat}\rm{Mo}}(\gamma,x\rm np)^{95\rm m,g}$Nb photonuclear reaction was studied using the electron beam from the NSC KIPT linear accelerator LUE-40. Experiment was performed using the activation and off-line $\gamma$-ray spectrometric technique. The experimental isomeric yield ratio $d(E_{\rm{\gamma max}}) = Y_{\rm m}(E_{\rm{\gamma max}}) / Y_{\rm g}(E_{\rm{\gamma max}})$  was determined for the reaction products $^{95\rm m,g}\rm{Nb}$ at the end-point bremsstrahlung energy $E_{\rm{\gamma max}}$ range of 35--95 MeV. The obtained values of $d(E_{\rm{\gamma max}})$ are in satisfactory agreement with the results of other authors and extend the range of previously known data. The theoretical values of the yields $Y_{\rm m,g}(E_{\rm{\gamma max}})$ and the isomeric yield ratio $d(E_{\rm{\gamma max}})$ for the isomeric pair $^{95\rm m,g}\rm{Nb}$ from the ${^{\rm nat}\rm{Mo}}(\gamma,x\rm np)$ reaction were calculated using the partial cross-sections  $\sigma(E)$ from the TALYS1.95 code for six different level density models $LD$. The comparison showed a noticeable excess (more than 3.85 times) of the experimental isomeric yield ratio over all theoretical estimates. At the investigated range of $E_{\rm{\gamma max}}$ the theoretical dependence of $d(E_{\rm{\gamma max}})$ on energy was confirmed -- the isomeric yield ratio smoothly decreases with increasing energy.
\end{abstract}


\maketitle


\section{Introduction}
\label{intro}

Photonuclear reactions are accompanied by the emission of a nucleon or a group of nucleons from a compound nucleus. This leads to an excited state of the final nucleus, the discharge of excitation energy of which occurs in a time of $10^{-12}$--$10^{-17}$ sec. However, in some cases, at low energy of the excitation level and a high degree of forbidden transition, long-lived excited states of atomic nuclei occur. Such states are called isomeric or metastable states, and their half-lives can vary from nanoseconds to many years \cite{0}.

Nuclei with isomeric (m) and unstable ground (g) states are of particular interest, since they allow to study of the metastable state population of this nucleus relative to its ground state, i.e. obtain the isomeric ratio  of the reaction products. This characteristic is defined as the ratio of the cross-sections for the formation of the reaction product in the metastable $\sigma_{\rm{m}}(E)$ and in the ground  $\sigma_{\rm{g}}(E)$ states: $d(E) = \sigma_{\rm{m}}(E)/\sigma_{\rm{g}}(E)$. The $d(E)$ values in the literature can be found as the ratio of the cross-sections for the formation of a product nucleus in the high-spin state (as a rule, this is a metastable state) $\sigma_{\rm{H}}(E)$ to the cross-section for the low-spin state $\sigma_{\rm{L}}(E)$: $d(E) = \sigma_{\rm{H}}(E)/\sigma_{\rm{L}}(E)$ \cite{1a,2a,3a}. In the case of research with the use of bremsstrahlung flux, this ratio $d(E_{\gamma\rm max})$ is expressed through the bremsstrahlung flux-averaged cross-sections or yields ($\langle{\sigma(E_{\rm{\gamma max}})}\rangle$ or $Y(E_{\gamma\rm max})$) of the reactions under study \cite{3a,4a,5a}.

Data on isomeric ratios of reaction products make it possible to investigate issues related to nuclear reactions and nuclear structure, such as the spin dependence of the nuclear level density, angular momentum transfer, nucleon pairing, shell effects, refine the theory of gamma transitions, and test theoretical models of the nucleus \cite{1,2,3,4,5}. The study of isomeric ratios using photonuclear reactions has an advantage since the $\gamma$-quantum introduces a small angular momentum and does not change the nucleon composition of the compound nucleus.

The studies in the energy range above giant dipole resonance (GDR) and before the pion production threshold (30--145 MeV) are of interest, because the mechanism of the nuclear reaction changes here: from dominance of GDR to dominance of the quasideuteron mechanism \cite{15a}. However, in this energy range, there is still a lack of experimental data on the cross-sections of multiparticle photonuclear reactions and isomeric ratios of the products of a nuclear reaction \cite{5b,6b}. This complicates the analysis of the mechanism of the nuclear reaction, as well as the systematization and analysis of the dependences of these quantities on various characteristics of the nucleus.

Experiments on the photodisintegration of stable isotopes of the Mo nucleus and determination of the isomeric ratio values $d(E_{\rm{\gamma max}})$ of the nuclide products $^{95\rm m,g}\rm{Nb}$ from the ${^{\rm nat}\rm{Mo}}(\gamma,x\rm np)$ reaction were performed in works \cite{15,16,17,18,19,20,21,22}, using beams of bremsstrahlung photons and the residual $\gamma$-activity method.

In the energy region of GDR, the formation of the isomeric pair $^{95\rm m,g}\rm{Nb}$ in the photonuclear reaction on ${^{\rm nat}\rm{Mo}}$ has been studied in works \cite{15,16,17,18,19}. Thus, in \cite{15,16} the experimental result for 
$d(E_{\rm{\gamma max}}) = Y_{\rm m}(E_{\rm{\gamma max}}) / (Y_{\rm g}(E_{\rm{\gamma max}}) + Y_{\rm m}(E_{\rm{\gamma max}}))$
was obtained at an energy of 30 MeV. The authors of \cite{17} defined the values of $d(E_{\rm{\gamma max}})$ as the ratio of $Y_{\rm m}(E_{\rm{\gamma max}}) / Y_{\rm g}(E_{\rm{\gamma max}})$ at energies of 25 and 30 MeV. In \cite{18,19} the range of investigated energies was extended to 14--24 MeV and the isomeric ratios of the reaction products of the $^{\rm nat}$Mo$(\gamma,x\rm np)^{95\rm m,g}$Nb reaction were also obtained by the $\gamma$-activation method. However, the results of \cite{18,19} contradict each other. It should be noted in works \cite{15,16,17,18,19} in obtained experimental results the main role in the formation of the isomeric pair $^{95\rm m,g}$Nb on the $^{\rm nat}$Mo plays the isotope $^{96}$Mo due to the reaction on the $^{97}$Mo nucleus has the higher threshold energy. 

At an intermediate energy region the 
 $^{\rm nat}$Mo$(\gamma,x\rm np)^{95\rm m,g}$Nb reaction was investigated in works \cite{20,21,22}. In \cite{20} the values of $d(E_{\rm{\gamma max}}) = Y_{\rm H}(E_{\rm{\gamma max}}) / Y_{\rm L}(E_{\rm{\gamma max}})$
 were obtained based on $(\gamma,\rm p)$, $(\gamma,\rm np)$, and $(\gamma,\rm 2np)$ reactions, which was the first measurement with $^{\rm nat}$Mo targets at $E_{\rm{\gamma max}}$ = 50, 60, 70 MeV. In \cite{21} authors extended the range of the study up to 45--70 MeV and attempted to compare the independent isomeric yield ratios of $^{95\rm m,g}$Nb for different reactions: $^{\rm nat}$Mo$(\gamma,x\rm np)$, $^{\rm nat}$Mo$({\rm p},\alpha x \rm n)$, and $^{\rm nat}$Zr$({\rm p},x \rm n)$.

In work \cite{22} the experimentally and theoretically studying the photodisintegration of molybdenum isotopes were performed. The yields of various photonuclear reactions on stable molybdenum isotopes were determined by the $\gamma$-activation method for the end-point bremsstrahlung energies of 19.5, 29.1, and 67.7 MeV. For the energy of 67.7 MeV, authors were able to determine the yields $Y_{\rm m}(E_{\rm{\gamma max}})$ and $Y_{\rm g}(E_{\rm{\gamma max}}) $, respectively, for the formation of the $^{95\rm m,g}$Nb nucleus in the isomeric and ground states, whose ratio is 5.5/5.1.

The experimental results from \cite{20,21,22} agree with each other, while the data from \cite{15,16,17,18,19} obtained in different representations of the value $d(E_{\rm{\gamma max}})$  contradict each other. This does not allow us to estimate the energy dependence of $d(E_{\rm{\gamma max}})$ in the wide region $E_{\rm{\gamma max}}$ = 14--70~MeV.

The present work is devoted to the study of the formation of the isomeric pair $^{95\rm m,g}$Nb in photonuclear reactions on $^{\rm nat}$Mo at the end-point bremsstrahlung energy range of $E_{\rm{\gamma max}}$ = 35--95 MeV. This will extend the range of previously known experimental isomeric yield ratios $d(E_{\rm{\gamma max}})$ and improve the reliability of these data.

\section{Experimental setup and procedure}
\label{sec:1}

An experimental study of the  $^{95\rm m,g}$Nb isomeric pair formation
 in the photonuclear reactions on $^{\rm nat}$Mo was performed using the electron beam of the linac LUE-40 RDC "Accelerator" NSC KIPT with the activation and off-line $\gamma$-ray spectrometric technique. The experimental procedure is described in detail, for example, in \cite{23,24,25,26}.

\begin{figure}[b]
	\resizebox{0.49\textwidth}{!}{%
		\includegraphics{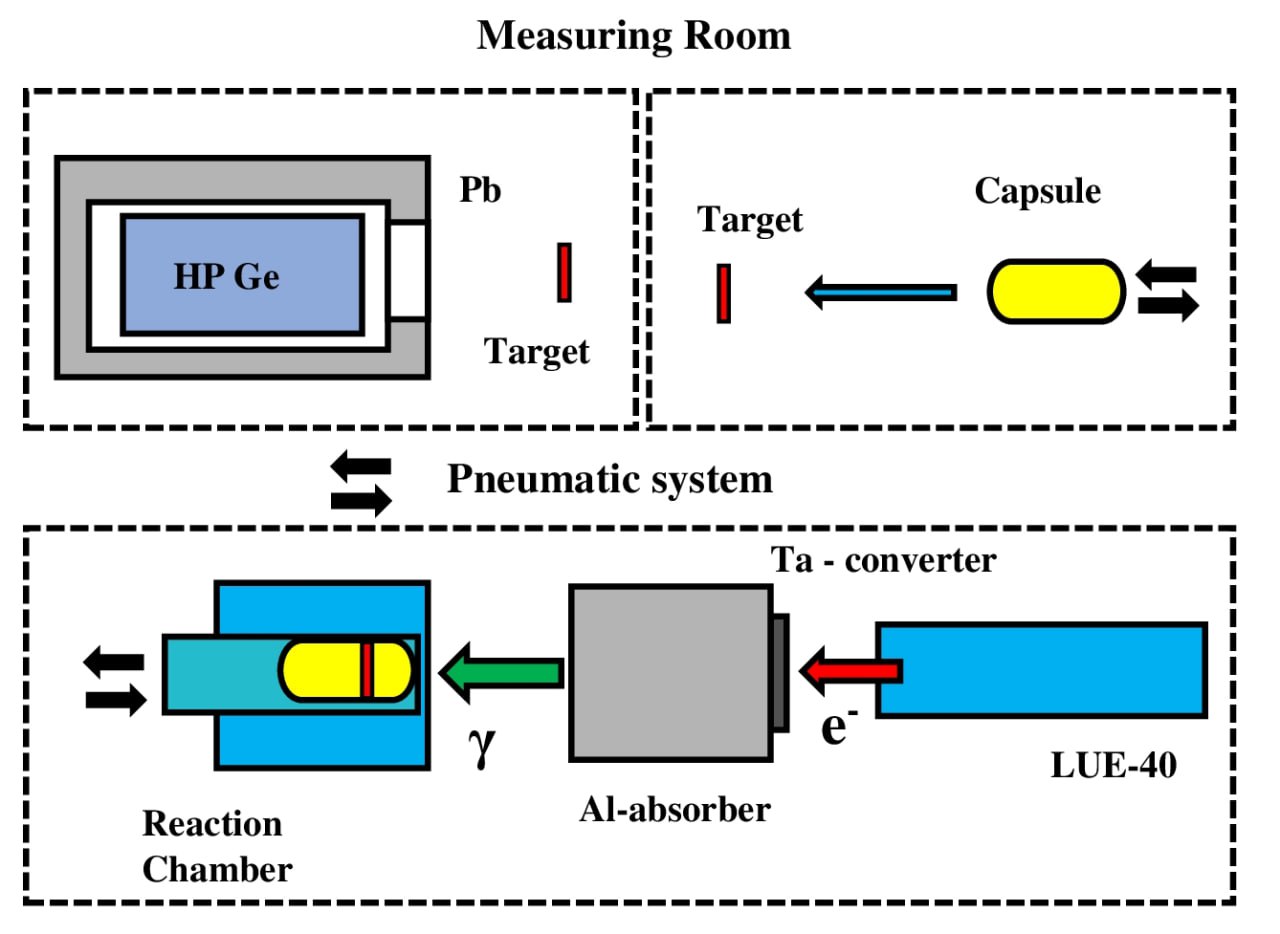}}
	\caption{Schematic block diagram of the experiment. The upper part shows the measurement room, where the irradiated target is extracted from the capsule and is arranged before the HPGe detector for induced $\gamma$-activity measurements. The lower part shows the linac LUE-40, the Ta converter, the Al absorber, and the exposure reaction chamber.}
	\label{fig1}
\end{figure}

The experimental complex for the study of photonuclear reactions is presented as a block diagram in Fig.~\ref{fig1}. The linac LUE-40 provides an electron beam with an average current $I_{\rm e} \approx 3~\mu \rm A$ and an energy spectrum with full width at half maximum (FWHM) $\Delta E_e/E_e \approx  1$\%. The range of initial energies of electrons $E_{\rm e}$ = 35--95 MeV. A detailed description and parameters of the LUE-40 linac are given in works \cite{27,28,29,30}.

On the axis of the electron beam, there are a converter, an absorber, and a reaction chamber. The converter, made of natural tantalum, is a $20\times20$~mm plate with thickness $l$ = 1.05~mm, and is attached to an aluminum absorber, shaped as a cylinder, with dimensions  $\oslash 100$~mm and a thickness of 150~mm. The thickness of the aluminum absorber was calculated to clean the beam of $\gamma$-quanta from electrons with energies up to 100 MeV.

In this experiment, targets were made of natural molybdenum, which were thin discs with a diameter of 8~mm and a thickness of $\approx$0.11 mm, which corresponded to a mass of $\approx$57--60~mg. Natural molybdenum consists of 7 stable isotopes, with isotope abundance, \%: $^{92}$Mo -- 14.84, $^{94}$Mo -- 9.25, $^{95}$Mo -- 15.92, $^{96}$Mo -- 16.68, $^{97}$Mo -- 9.55, $^{98}$Mo -- 24.13, $^{100}$Mo -- 9.63 (according to \cite{31,32}).

\begin{figure*}[]
	\begin{minipage}[h]{0.99\linewidth}
		{\includegraphics[width=1\linewidth]{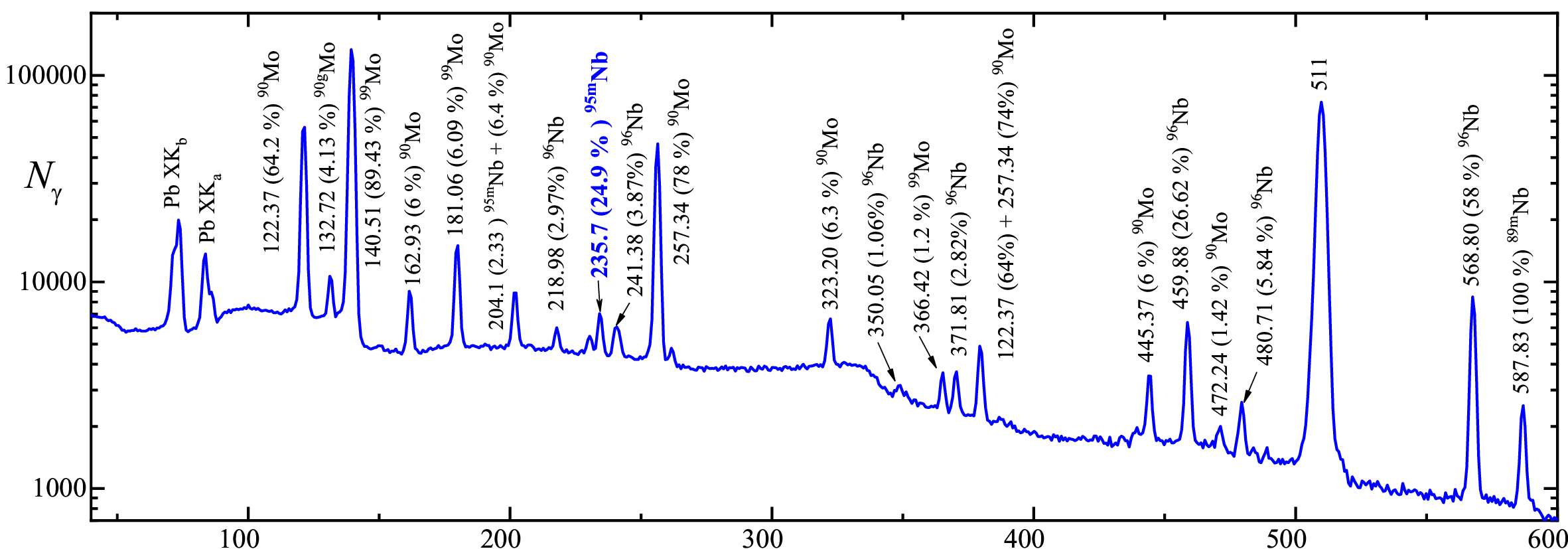}} \\
	\end{minipage}
	\vfill
	\begin{minipage}[h]{0.99\linewidth}
		{\includegraphics[width=1\linewidth]{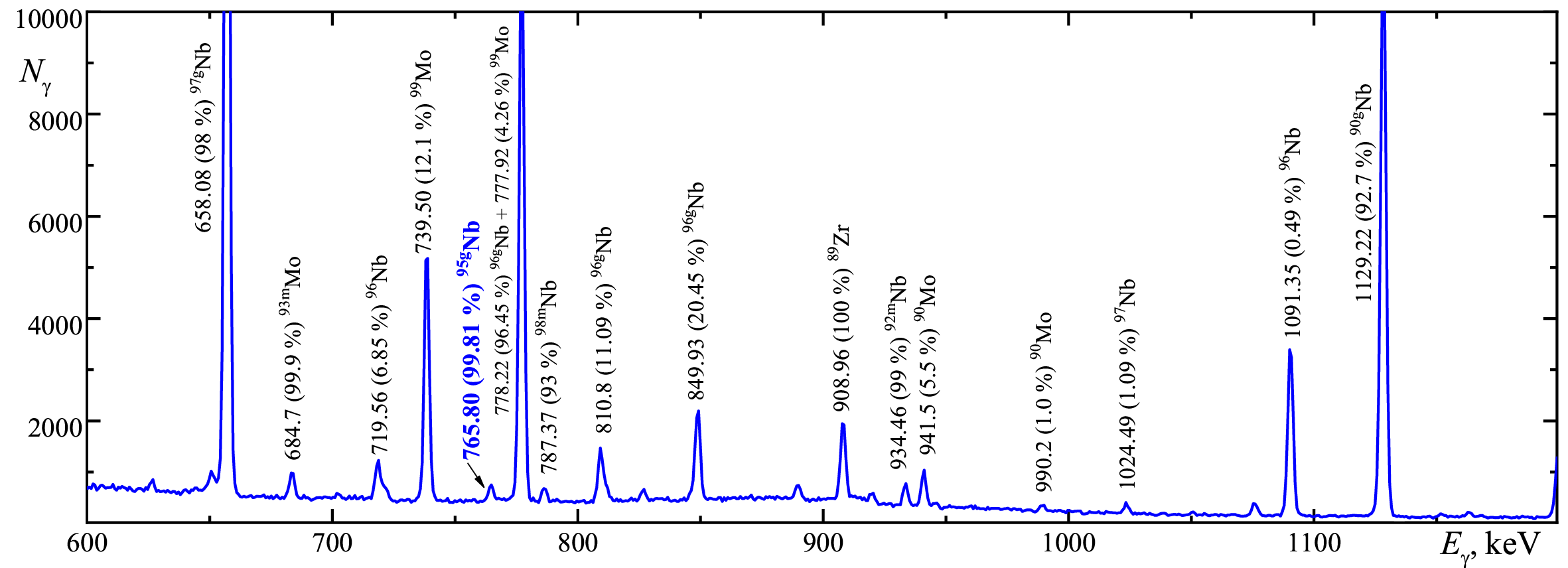}} \\
	\end{minipage}
	\caption{Fragments of $\gamma$-ray spectrum in the energy ranges $40 \leq E_{\rm \gamma} \leq 600$~keV and $600 \leq E_{\rm \gamma} \leq 1200$~keV from the $^{\rm nat}$Mo target of mass 57.862~mg after irradiation of the bremsstrahlung $\gamma$-flux at the end-point bremsstrahlung energy $E_{\rm{\gamma max}}$ = 92.50~MeV. The irradiation $t_{\rm irr}$ and measurement $t_{\rm meas}$ times were both 3600~s.}
	\label{fig2}
\end{figure*}

The targets were placed in an aluminum capsule and using a pneumatic transport system were taken to the reaction chamber for irradiation and back to the measurement room to record the induced $\gamma$-activity of reaction products in the target substance.

The $\gamma$-quanta of the reaction products were detected using a Canberra GC-2018 semiconductor HPGe detector. At energy $E_\gamma$ = 1332~keV its efficiency was 20\% relative to the NaI(Tl) scintillator with dimensions 3 inches in diameter and 3 inches in thickness. The resolution FWHM is 1.8~keV for energy $E_\gamma$ = 1332 keV, and 0.8~keV for $E_\gamma$ = 122 keV. The dead time for $\gamma$-quanta detection varied between 0.1 and 5\%. The absolute detection efficiency $\varepsilon (E_\gamma)$ for $\gamma$-quanta of different energies was obtained using a standard set of $\gamma$-rays sources: $^{22}$Na, $^{60}$Co, $^{133}$Ba, $^{137}$Cs, $^{152}$Eu, $^{241}$Am. 

The electron bremsstrahlung spectra were calculated using the open-source software code GEANT4.9.2, PhysList G4LowEnergy \cite{32}. The real geometry of the experiment was used in calculations as well as the space and energy distributions of the electron beam were taken into account. 

Additionally, the bremsstrahlung flux was monitored by the yield of the $^{100}\rm{Mo}(\gamma,n)^{99}\rm{Mo}$ reaction (the half-life of the $^{99}\rm{Mo}$ nucleus is $T_{1/2} = 65.94 \pm 0.01$ h). For this purpose,   the $\gamma$-line with the energy of $E_\gamma$ = 739.5~keV and the absolute intensity $I_\gamma = 12.13 \pm 0.12$\% [33] was used (see for more detail \cite{23,24,25,26}).

The $\gamma$-radiation spectrum of a $^{\rm nat}$Mo target irradiated by a beam of bremsstrahlung  $\gamma$-quanta with the high end-point energy has a complex pattern. There are emission  $\gamma$-lines of nuclei product of the $^{\rm nat}$Mo$(\gamma,x{\rm n}y{\rm p})$
 reactions located on a background substrate, which is formed as a result of Compton scattering of photons. As an example, $\gamma$-radiation spectrum of
  a $^{\rm nat}$Mo target with a mass of 57.862~mg after irradiation with $E_{\rm{\gamma max}}$ = 92.50~MeV is shown in Fig.~\ref{fig2}.

\section{Theoretical calculation}
\label{sec:2}

The $^{95}$Nb nucleus in the isomeric and ground states can be formed in photonuclear reactions  $^{\rm nat}$Mo$(\gamma,x{\rm np})$. Natural molybdenum consists of 7 stable isotopes, but only four isotopes contribute to the formation of the $^{95}$Nb nucleus; respectively, there are four reactions with thresholds: 

$^{96}$Mo$(\gamma,\rm p)^{95\rm g}$Nb -- $E_{\rm thr}$ = 9.30 MeV;

$^{97}$Mo$(\gamma,\rm np)^{95\rm g}$Nb -- $E_{\rm thr}$ = 16.12 MeV;

$^{98}$Mo$(\gamma,\rm 2np)^{95\rm g}$Nb -- $E_{\rm thr}$ = 24.76 MeV;

$^{100}$Mo$(\gamma,\rm 4np)^{95\rm g}$Nb -- $E_{\rm thr}$ = 38.98 MeV. 

The thresholds for the formation of the  $^{95\rm m}$Nb nucleus in the metastable state are higher than in the ground state at an excitation energy of 235.7 keV.

The calculation of theoretical cross-sections $\sigma(E)$ of studied reactions for monochromatic photons was performed using the TALYS1.95 code \cite{31}, which is installed on Linux Ubuntu-20.04. The calculations were performed for different level density models $LD$ 1-6. There are three phenomenological level density models and three options for microscopic level densities:

$LD 1$: Constant temperature + Fermi gas model, introduced by Gilbert and Cameron \cite{Gilbert}.

$LD 2$: Back-shifted Fermi gas model \cite{Back-shifted}. 

$LD 3$: Generalized superfluid model (GSM) \cite{Ignatyuk1,Ignatyuk2}. 

$LD 4$: Microscopic level densities (Skyrme force) from Goriely’s tables \cite{Goriely}. 

$LD 5$: Microscopic level densities (Skyrme force) from Hilaire’s combinatorial tables \cite{Hilaire}. 

$LD 6$: Microscopic level densities based on temperature-dependent Hartree-
Fock-Bogoliubov calculations using the Gogny force from Hilaire’s combinatorial tables \cite{Gorny}.

\begin{figure}[t]
	\resizebox{0.49\textwidth}{!}{%
		\includegraphics{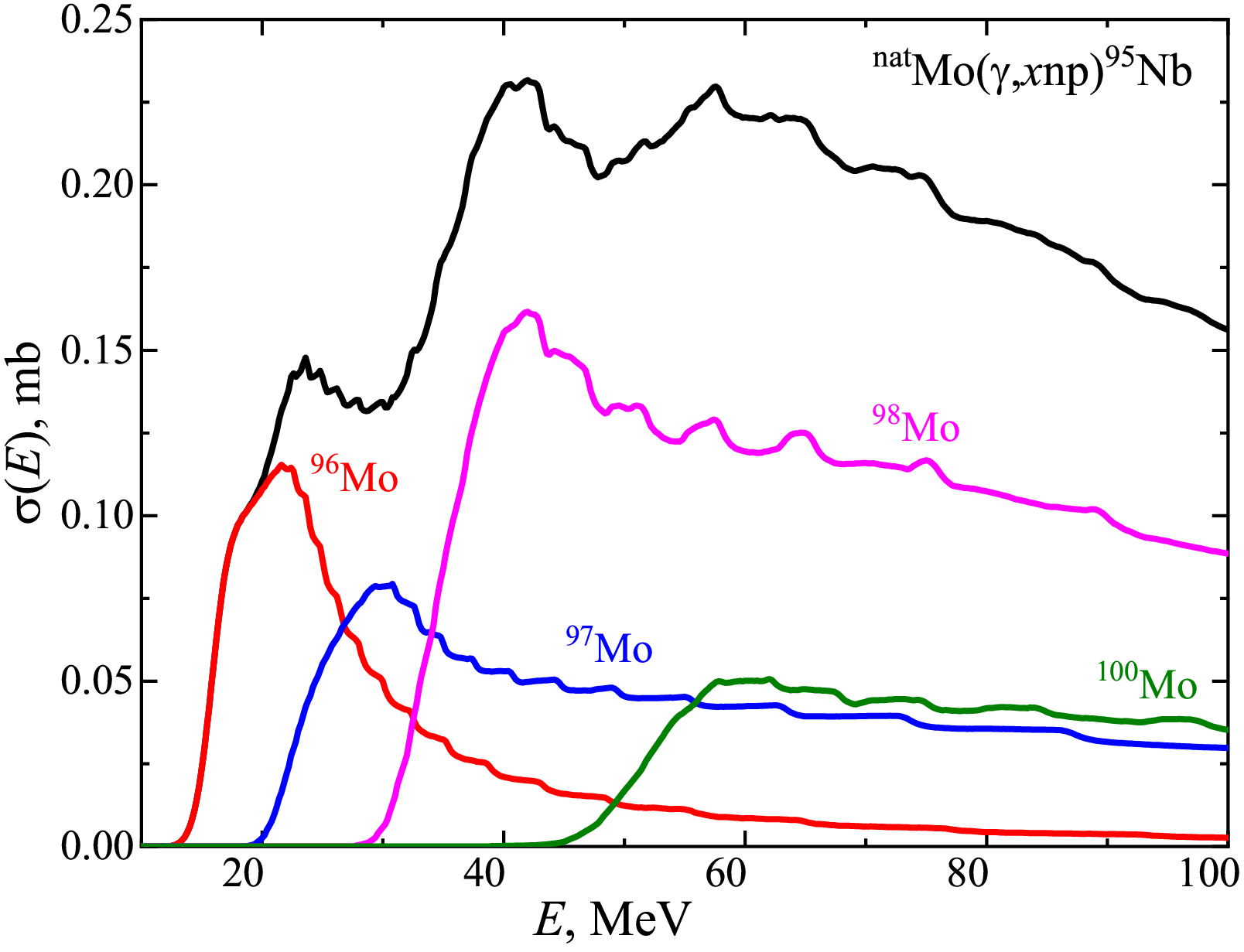}}
	\caption{Theoretical total (metastable + ground) cross-sections $\sigma(E)$ for the formation of the $^{95}$Nb nucleus on stable isotopes of molybdenum $^{96,97,98,100}$Mo (color curves) and on $^{\rm nat}$Mo (black curve). The calculations were performed for the level density model $LD$1.}
	\label{fig3}
\end{figure}

Fig.~\ref{fig3} shows the total (metastable + ground) cross-sections $\sigma(E)$ for the formation of the $^{95}$Nb nucleus on 4 stable isotopes of molybdenum (96, 97, 98, 100) calculated in the TALYS1.95 code, $LD$1. The cross-sections are given taking into account the abundance of isotopes. As can be seen, in the energy range above 35~MeV, the cross-section of the $^{98}$Mo$(\gamma,\rm 2np)^{95}$Nb reaction has the highest values. This is due both to the $^{98}$Mo isotopic abundance of 24.13\% and to insignificant differences in the cross-sections for the formation of the $^{95}$Nb nucleus on different molybdenum isotopes.

\begin{figure}[t]
	\resizebox{0.49\textwidth}{!}{%
		\includegraphics{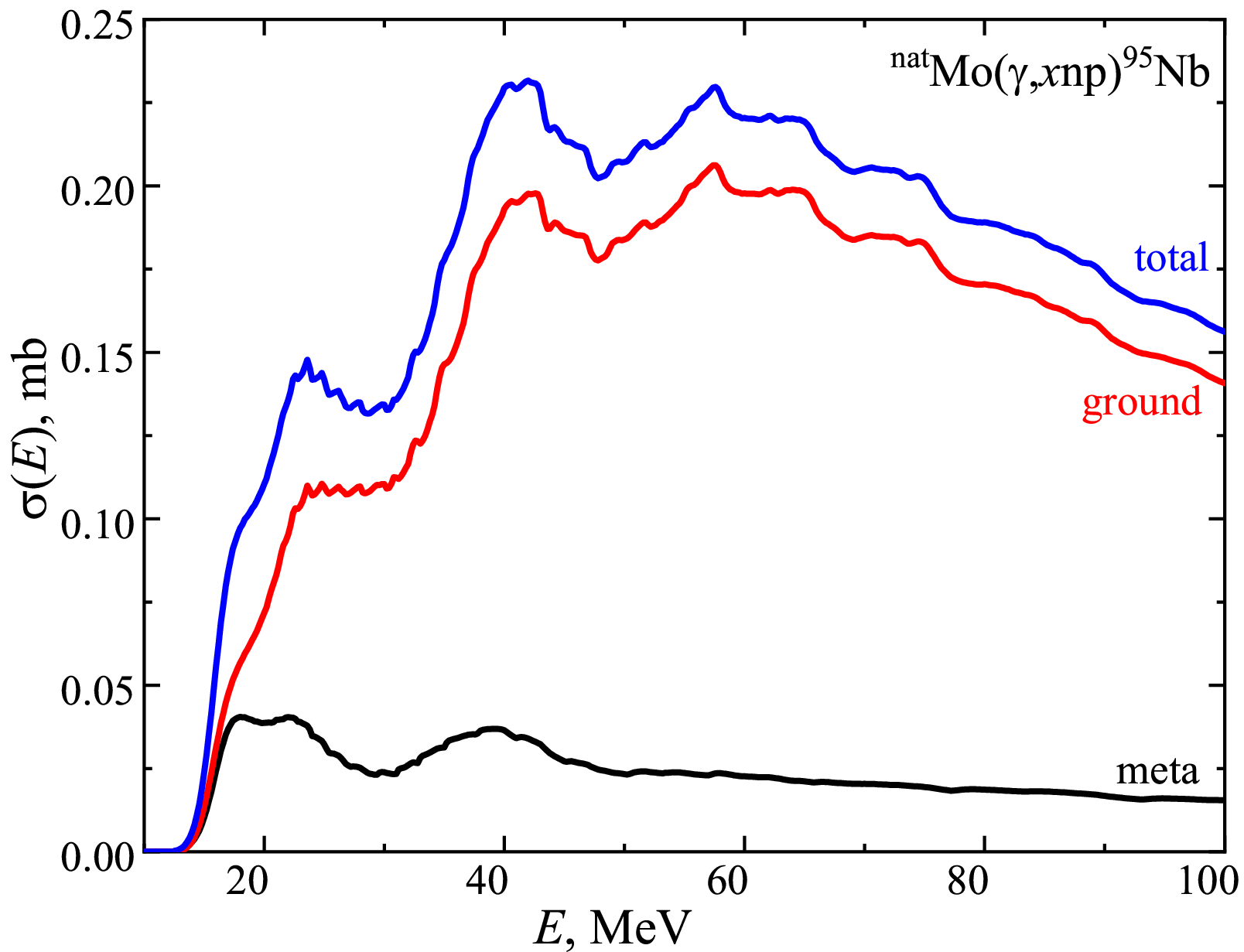}}
	\caption{Theoretical cross-sections $\sigma(E)$ of the $^{\rm nat}$Mo$(\gamma,x\rm np)^{95\rm m,g}$Nb reactions. The calculations were performed for the level density model $LD$1.}
	\label{fig4}
\end{figure}

Fig.~\ref{fig4} shows the cross-sections $\sigma(E)$ for the formation of the $^{95}$Nb nucleus in the metastable and ground states in the $^{\rm nat}$Mo$(\gamma,x\rm np)$ reactions and the total cross-section calculated in the TALYS1.95 code, $LD$1. As can be seen from figure, the contribution of the metastable state to the total cross-section does not exceed  20\% at energies above 30 MeV.

Using the theoretical cross-sections $\sigma(E)$, one can obtain the reaction yield, which is determined by the formula:

\begin{equation}\label{form1}
	Y(E_{\rm{\gamma max}}) = N_n \int\limits_{E_{\rm{thr}}}^{E_{\rm{\gamma max}}}\sigma(E)\cdot W(E,E_{\rm{\gamma max}})dE,
\end{equation}
where $N_n$ is the number of atoms of the element under study, $W(E,E_{\rm{\gamma max}})$  is the bremsstrahlung $\gamma$-flux, $E_{\rm thr}$ -- an energy of the reaction threshold, $E_{\rm{\gamma max}}$ -- the end-point bremsstrahlung energy. 

To estimate the $i$-th reaction contribution in the total production of a studied nuclide (for example, the $^{96}$Mo$(\gamma,p)$ reaction in the production of the $^{95}$Nb nucleus on $^{\rm nat}$Mo), the relative reaction yield $Y_{\rm i}(E_{\rm{\gamma max}})$ were used. To calculate the $Y_{\rm i}(E_{\rm{\gamma max}})$ we used the cross-sections from the TALYS1.95 code and expression:

     \begin{equation}\label{form2}
	Y_{\rm i}(E_{\rm{\gamma max}}) = \frac
	{A_i \int\limits_{E^i_{\rm{thr}}}^{E_{\rm{\gamma max}}}\sigma_i(E)\cdot W(E,E_{\rm{\gamma max}})dE}
	{\sum^4_{k=1} A_k \int\limits_{E^k_{\rm{thr}}}^{E_{\rm{\gamma max}}}\sigma_k(E)\cdot W(E,E_{\rm{\gamma max}})dE},
\end{equation}
where $\sigma_k(E)$ is the cross-section for the formation of the $^{95}$Nb nucleus on the $k$-th isotope with isotopic abundance $A_k$. Summation over $k$ was carried out for 4 stable molybdenum isotopes $^{96,97,98,100}$Mo.

\begin{figure}[t]
	\resizebox{0.49\textwidth}{!}{%
		\includegraphics{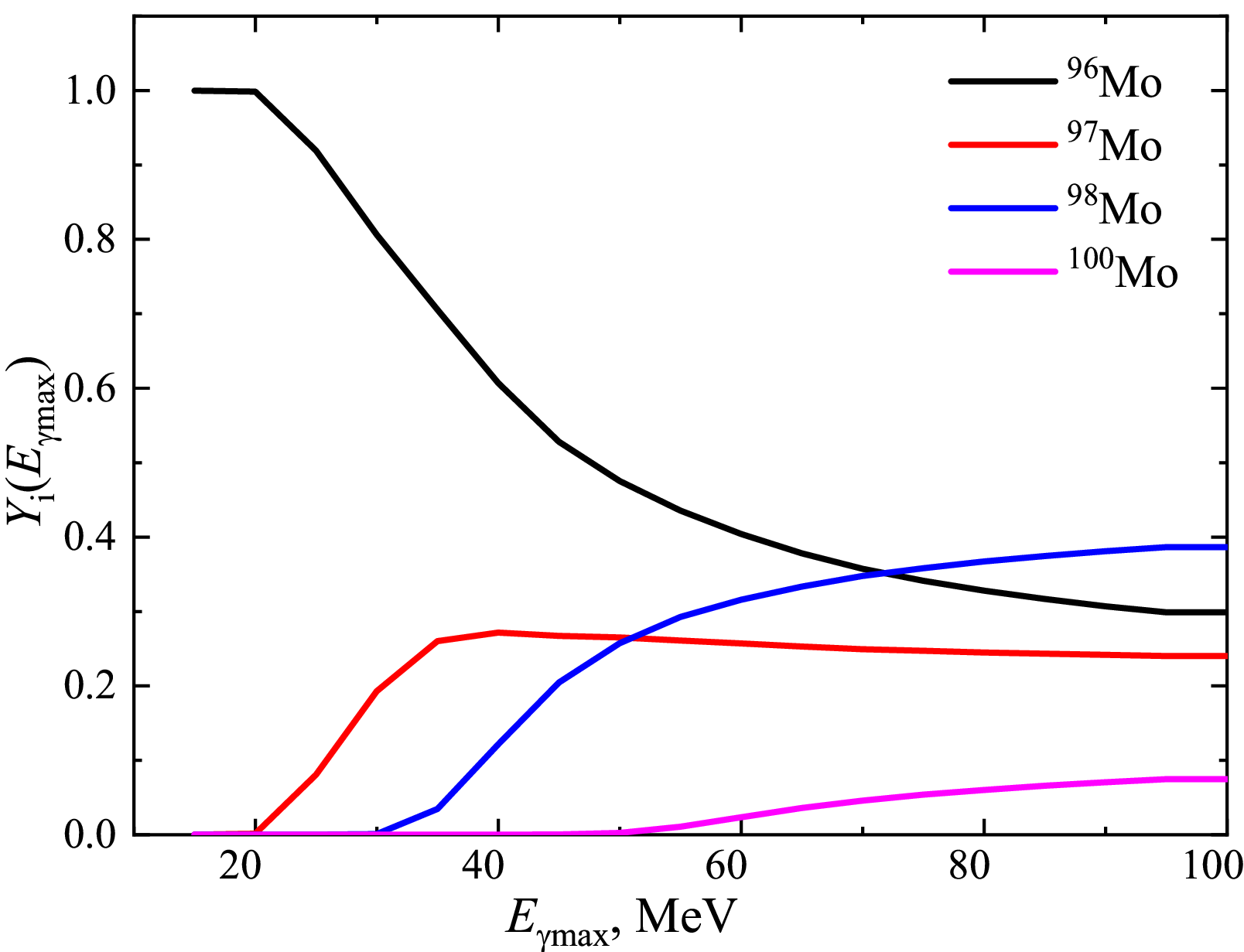}}
	\caption{Theoretical reaction yields of the $^{95}$Nb formation on various molybdenum isotopes relative to the total yield on $^{\rm nat}$Mo according to Eq.~\ref{form2}. The sum of the relative yield is equal to 1.0. The calculations were performed for the level density model $LD$1. }
	\label{fig5}
\end{figure}

Fig.~\ref{fig5} shows the contributions of the yields of various isotopes to the total yield of the $^{\rm nat}$Mo$(\gamma,x{\rm np})^{95}$Nb reaction according to Eq.~\ref{form2} with the use  cross-sections from TALYS1.95 code, $LD$1. The contribution of the reaction yield on a given isotope is determined by the cross-section, the reaction threshold, and the isotope abundance.

As a rule, in the presence of several isotopes, there is one whose contribution to the reaction yield dominates ($>$ 90\%),  for example, as shown in \cite{34,35}. In the case of the reaction under study at energies up to 20 MeV, the contribution of $^{96}$Mo is 100\%. But at the end-point bremsstrahlung energy above 30 MeV, it is difficult to determine the dominant reaction. Thus, for natural molybdenum, it is necessary to take into account the contribution of all stable isotopes to the yield of the $^{\rm nat}$Mo$(\gamma,x\rm np)^{95}$Nb reaction.

\begin{figure}[t]
	\resizebox{0.49\textwidth}{!}{%
		\includegraphics{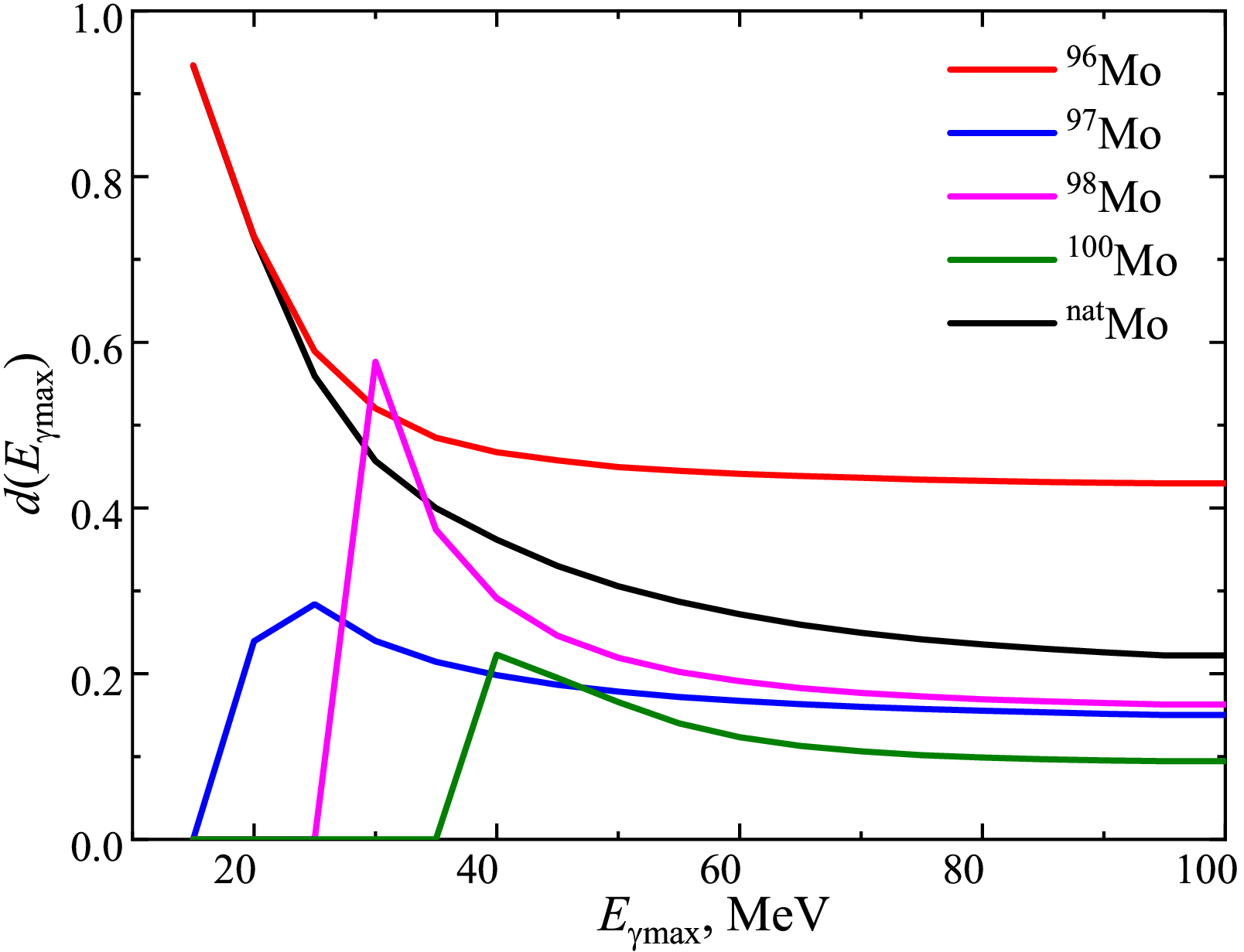}}
	\caption{Theoretical isomeric yield ratios $d(E_{\rm{\gamma max}})$ for the reaction products $^{95\rm m,g}$Nb on different molybdenum isotopes and $^{\rm nat}$Mo. The calculations were performed for the level density model $LD$1.}
	\label{fig6}
\end{figure}

\begin{figure}[t]
	\resizebox{0.49\textwidth}{!}{%
		\includegraphics{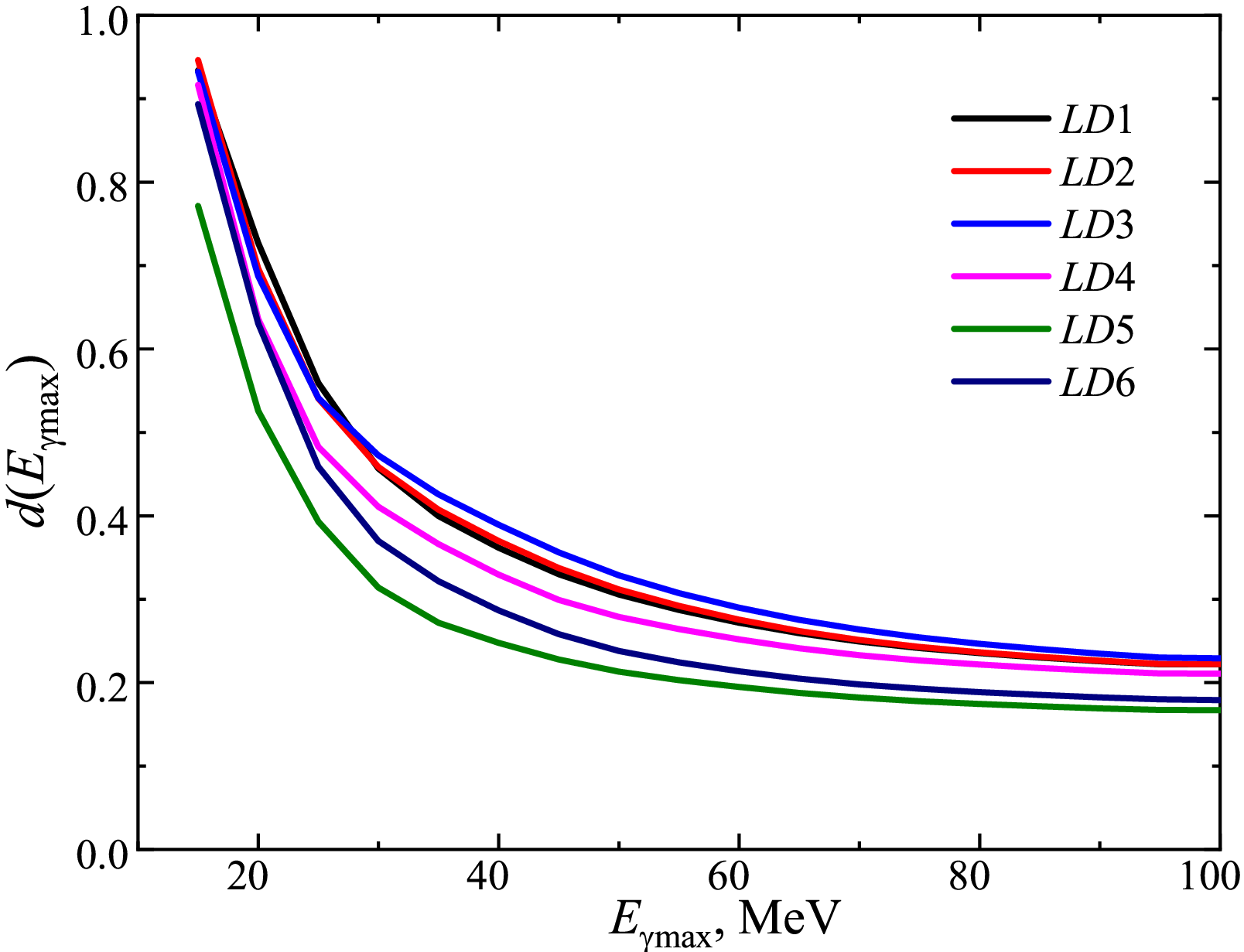}}
	\caption{Theoretical isomeric yield ratios $d(E_{\rm{\gamma max}})$ for the reaction products from the $^{\rm nat}$Mo$(\gamma,x\rm np)^{95\rm m,g}$Nb reaction for different level density models $LD$ 1-6.}
	\label{fig6a}
\end{figure}

Figures~\ref{fig3}--\ref{fig5} show the results of calculations only for the level density model $LD$1. This was done so as not to clutter up the figures with many curves.

 In this work, the values of the isomeric yield ratio are calculated as the ratio of the yield $Y_{\rm m}(E_{\rm{\gamma max}})$  of the formation of the nucleus in the metastable state to the yield $Y_{\rm g}(E_{\rm{\gamma max}})$ of the formation of the nucleus in the ground state:
\begin{equation}\label{form3}
	d(E_{\rm{\gamma max}}) =
	Y_{\rm m}(E_{\rm{\gamma max}}) / Y_{\rm g}(E_{\rm{\gamma max}}).
\end{equation}

Fig.~\ref{fig6} shows the theoretical prediction of the isomeric yield ratio $d(E_{\rm{\gamma max}})$ calculated using the cross-sections from the TALYS1.95 code, $LD$1 model. As can be seen, the calculated isomeric yield ratios differ for different isotopes 96--100. The value $d(E_{\rm{\gamma max}})$ calculated for natural Mo is between them.

 The theoretical estimations of the isomeric yield ratio
 	$d(E_{\rm{\gamma max}})$  were calculated using the cross-sections from the
 	TALYS1.95 code for six level density models $LD$ and
 	shown in Fig.~\ref{fig6a}. As can be seen, the theoretical isomeric
 	yield ratios differ on (57--37)\% at the range of $E_{\rm{\gamma max}}$ = 35--100 MeV. The $d(E_{\rm{\gamma max}})$
 	calculated for the $LD$3 model has the highest value at the energy
 	range of $E_{\rm{\gamma max}}$ = 30--100 MeV.

\section{Experimental results}
\label{sec:4}

\begin{figure}[t]
	\resizebox{0.49\textwidth}{!}{%
		\includegraphics{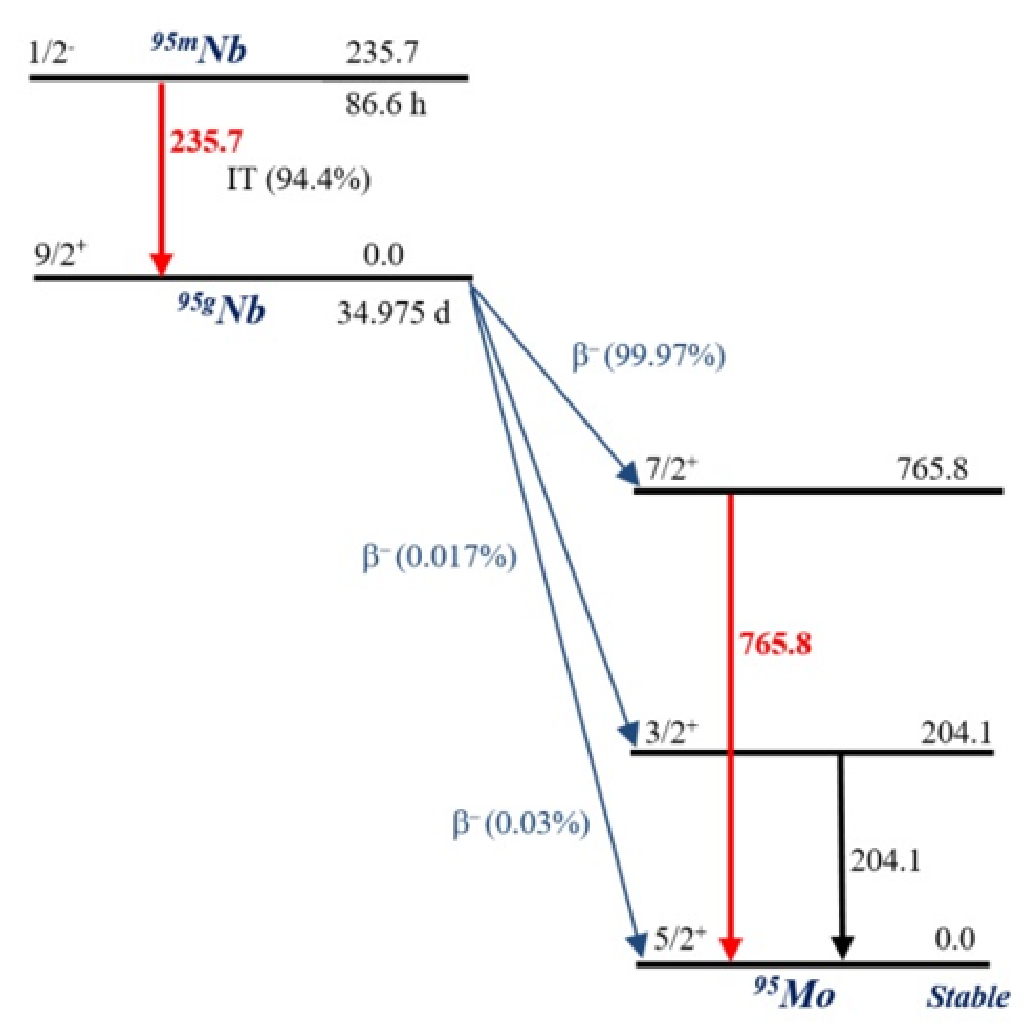}}
	\caption{Simplified representation of formation and decay scheme of the isomeric pair $^{95\rm m,g}$Nb. The nuclear level energies are in keV. The red color shows the emission $\gamma$-lines that were used in this work for the analysis. }
	\label{fig7}
\end{figure}

A simplified diagram of the decay of the niobium nucleus from the ground $^{95\rm g}$Nb and metastable $^{95\rm m}$Nb states is shown in Fig.~\ref{fig7}. Nuclear spectroscopic data of the reactions $^{\rm nat}$Mo$(\gamma,x\rm np)^{95\rm m,g}$Nb are presented according to \cite{33}. 

The isomeric state $^{95\rm m}$Nb (low-spin state, $J^\pi = 1/2^{-}$) with a half-life $T_{1/2}$ of $86.6 \pm 0.08$ h decays to the unstable ground state $^{95\rm g}$Nb (high-spin state, $J^\pi = 9/2^{+}$) by emitting $\gamma$-quanta with the energy of 235.7~keV through an internal transition with a branching ratio $p$ of $94.4 \pm 0.6$\%. Meanwhile, 5.6\% of the isomeric state decays to the various energy levels of stable $^{95}$Mo by a $\beta^-$-process. The unstable ground state $^{95\rm g}$Nb with a half-life $T_{1/2}$ of $34.975 \pm 0.007$ d decays to the 765.8 keV energy level of $^{95}$Mo by a $\beta^-$-process. 

This decay pattern leads to the ground state radionuclide $^{95\rm g}$Nb can be formed in two ways, directly from the target nuclide and/or indirectly through the decay of the metastable radionuclide. 

As a result, to find the experimental values of the isomeric yield ratio $d(E_{\rm{\gamma max}})$, it is necessary to solve the system of equations describing the radioactive decay of the isomeric state and decay with accumulation for the ground state. The solution of such a system of equations is given in several works, for example, \cite{36,37,38,39}, in different analytical representations.

In this work to determine the experimental values of the isomeric yield ratio $d(E_{\rm{\gamma max}})$, the following expression was used according to \cite{36}: 

     \begin{equation}\begin{split}
     		\label{form4}
	d(E_{\rm{\gamma max}}) =
	Y_{\rm m}(E_{\rm{\gamma max}}) / Y_{\rm g}(E_{\rm{\gamma max}}) = \;\;\;\;\;\;\;\;\;\;\;\;\;\;\;\;\;\;\;\;\;\;\;\;\;\;\\
 \left[  \frac{\lambda_{\rm g} F_{\rm m}(t)}{\lambda_{\rm m} F_{\rm g}(t)}
	\left(\frac{\Delta A_{\rm g} I_{\rm m} \varepsilon_{\rm m}}{\Delta A_{\rm m} I_{\rm g} \varepsilon_{\rm g}} -
	p\frac{\lambda_{\rm g}}{\lambda_{\rm g} - \lambda_{\rm m}}\right) +
	p\frac{\lambda_{\rm m}}{\lambda_{\rm g} - \lambda_{\rm m}}
	\right]^{-1},
	\end{split}
\end{equation}
     \begin{equation}
     	F_{\rm m}(t) = (1-e^{-\lambda_{\rm m} t_{\rm irr}})e^{-\lambda_{\rm m} t_{\rm cool}}(1-e^{-\lambda_{\rm m} t_{\rm meas}}),
     \end{equation}
 \begin{equation}
 	F_{\rm g}(t) = (1-e^{-\lambda_{\rm g} t_{\rm irr}})e^{-\lambda_{\rm g} t_{\rm cool}}(1-e^{-\lambda_{\rm g} t_{\rm meas}}),
 \end{equation}
where $\lambda_{\rm g}$ and $\lambda_{\rm m}$ are the decay constants for the ground and isomeric states, respectively; $\Delta A_{\rm g}$ and $\Delta A_{\rm m}$ are the number of counts in the peaks at the energies of $\gamma$-quanta corresponding to the decays of the isomeric and ground states, respectively; $\varepsilon_{\rm g}$ and $I_{\rm g}$ ($\varepsilon_{\rm m}$ and $I_{\rm m}$) are the detector efficiency and the absolute intensity of a $\gamma$-quantum with an energy corresponding to the decay of the ground state (isomeric state); $p$ -- the branching ratio for the decay of the isomeric to the ground state (94.4\%); $t_{\rm irr}$, $t_{\rm cool}$ and $t_{\rm meas}$ are the irradiation time, cooling time, and measurement time, respectively.

In natural molybdenum targets, as a result of the $^{\rm nat}$Mo$(\gamma,x\rm n2p)$ reaction, the $^{95}$Zr nucleus can also be formed. In the decay scheme of $^{95}$Zr ($T_{1/2} = 65.02 \pm 0.05$~d) there is a
 $\gamma$-transition with the energy $E_\gamma$ = 235.7~keV and intensity $I_\gamma = 0.294 \pm 0.016$\%. The decay of the $^{95}$Zr nucleus can contribute to the observed value of $\Delta A\rm _m$. To take this
 contribution into account, calculations were performed in the TALYS1.95 code, level density model $LD$1. It was found that the estimated activity of $^{95}$Zr by $\gamma$-line with 235.7 keV is negligible. The contribution of $^{95}$Zr was also experimentally verified by the $\gamma$-lines corresponding to the decay of the $^{95}$Zr nucleus, namely, $E_\gamma$ = 724.2 keV ($I_\gamma = 44.17 \pm 0.13$\%) and $E_\gamma$ = 756.7 keV ($I_\gamma$ = 54\%). No such peaks were found in the measured spectra of the induced $\gamma$-activity of the targets. 

For the $\gamma$-quanta with an energy of 235.7~keV and the thicknesses of the molybdenum targets used, the self-absorption coefficients were calculated using the GEANT4.9.2 code. It was found that the value of the self-absorption coefficient did not exceed 0.8\%, which was taken into account when processing the experimental results.

\begin{figure}[t]
	\resizebox{0.49\textwidth}{!}{%
		\includegraphics{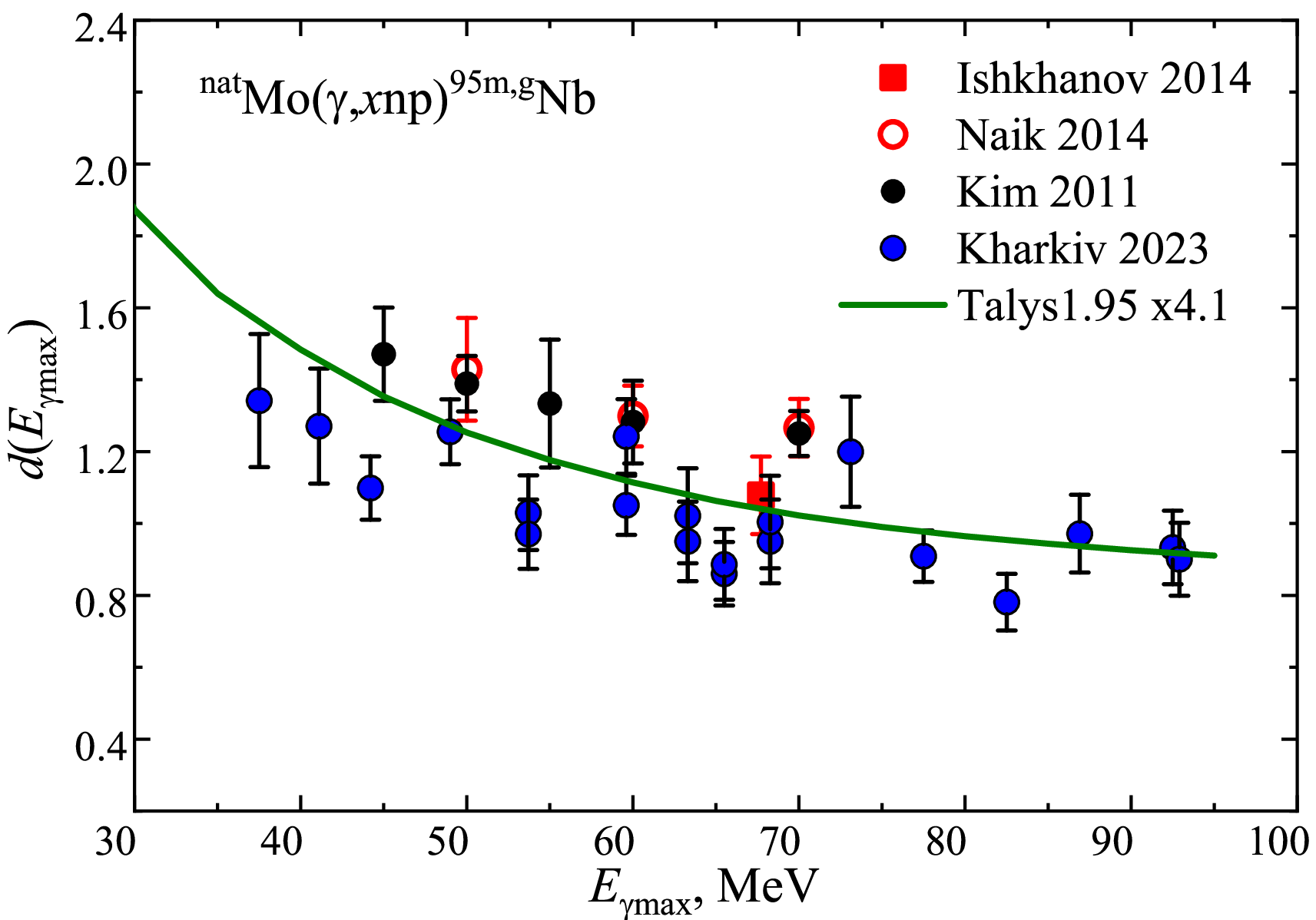}}
	\caption{Isomeric yield ratio $d(E_{\rm{\gamma max}})$ for the reaction products from the $^{\rm nat}$Mo$(\gamma,x\rm np)^{95\rm m,g}$Nb reaction. Experimental results of this work -- blue circles, data of other authors: square -- \cite{22}, red empty circles -- \cite{20}, black circles -- \cite{21}. The curve -- calculation in the code TALYS1.95, $LD$3, multiplied by a factor of 3.85. }
	\label{fig8}
\end{figure}

The experimental values of the isomeric yield ratio $d(E_{\rm{\gamma max}})$ of the nuclei products from the $^{\rm nat}$Mo$(\gamma,x\rm np)^{95\rm m,g}$Nb reaction were determined at the end-point bremsstrahlung energy of 35--95 MeV (see Fig.~\ref{fig8} and Table~\ref{tab1}).

The calculation of the experimental error of the values $d(E_{\rm{\gamma max}})$ was performed taking into account statistical and systematic errors, the description of which can be found, for example, in \cite{23,24}. The uncertainties related to the $d(E_{\rm{\gamma max}})$ were calculated from Eq.~\ref{form4} by using the error propagation principle, which indicates the maximum uncertainty of the measured values in Fig.~\ref{fig8} and Table~\ref{tab1}. 

 \begin{table}[h]
	\caption{\label{tab1} Isomeric yield ratio $d(E_{\rm{\gamma max}})$ for the nuclei products $^{95\rm m,g}\rm{Nb}$ from	the ${^{\rm nat}\rm{Mo}}(\gamma,x\rm np)$ reaction.}
	\centering
	\begin{tabular}{cc}
		\hline\hline\noalign{\smallskip}
		$E_{\rm{\gamma max}}$,~MeV & $\;\;\;\;\;\; d(E_{\rm{\gamma max}}) \pm \Delta d(E_{\rm{\gamma max}})$   \\ \noalign{\smallskip}\hline\noalign{\smallskip}	
		37.50 & 1.34 $\pm$ 0.19  \\ 
		41.10 & 1.27 $\pm$ 0.16  \\ 
		44.20 & 1.10 $\pm$ 0.09  \\ 
		49.00 & 1.25 $\pm$ 0.09  \\ 
		53.70 & 1.03 $\pm$ 0.10  \\ 
		53.70* & 0.97 $\pm$ 0.10  \\ 
		59.60 & 1.05 $\pm$ 0.08  \\ 
		59.60* & 1.24 $\pm$ 0.10  \\ 
		63.30 & 0.95 $\pm$ 0.11  \\ 
		63.30* & 1.02 $\pm$ 0.13  \\ 
		65.50 & 0.86 $\pm$ 0.09  \\ 
		65.50* & 0.89 $\pm$ 0.10  \\ 
		68.25 & 0.95 $\pm$ 0.11  \\ 
		68.25* & 1.00 $\pm$ 0.13  \\ 
		73.10 & 1.20 $\pm$ 0.15  \\ 
		77.50 & 0.91 $\pm$ 0.08  \\ 
		82.50 & 0.78 $\pm$ 0.08  \\ 
		86.90 & 0.97 $\pm$ 0.11  \\ 
		92.50 & 0.93 $\pm$ 0.10  \\ 
		92.90 & 0.90 $\pm$ 0.10  \\ \noalign{\smallskip}\hline\hline	  	
	\end{tabular} \\ 
	\footnotesize{* an additional series of measurements.}
\end{table}

\section{Discussion}
\label{Disc}

Two main representations of the isomeric ratio $d(E_{\rm{\gamma max}})$ are used in the literature. 
 One of them is defined as the ratio of the yields  $Y_{\rm m}(E_{\rm{\gamma max}})$ and $Y_{\rm g}(E_{\rm{\gamma max}})$ states: $d(E_{\rm{\gamma max}}) = Y_{\rm m}(E_{\rm{\gamma max}}) / Y_{\rm g}(E_{\rm{\gamma max}})$ (for example, see in \cite{17}). The $d(E_{\rm{\gamma max}})$ values also can be found as the ratio of the yield for the formation of a product nucleus in the high-spin state 
 $Y_{\rm H}(E_{\rm{\gamma max}})$ to the yield for the low-spin state $Y_{\rm L}(E_{\rm{\gamma max}})$: $d(E_{\rm{\gamma max}}) = Y_{\rm H}(E_{\rm{\gamma max}}) / Y_{\rm L}(E_{\rm{\gamma max}})$ \cite{20,21}. These values will be equal to each other if the nucleus in the isomeric state has a larger spin. However, the ground state of the $^{95}$Nb nucleus has a high spin $J^\pi = 9/2^{+}$, while the spin of the $^{95\rm m}$Nb isomeric state is $J^\pi = 1/2^{-}$. This means that the values of $d(E_{\rm{\gamma max}})$ described above are inversely proportional to each other and are presented differently in different works.

In our work the experimental isomeric yield ratios of the pair $^{95\rm m,g}$Nb were obtained by Eq.~\ref{form4}. A comparison of our results with the data of other authors \cite{20,21} shows satisfactory agreement within the experimental error. The estimate of the value of $d(E_{\rm{\gamma max}})$, obtained using the yield values from \cite{22}, is in agreement with both our results and the data from \cite{20,21}.

The range of changes in the experimental values $d(E_{\rm{\gamma max}})$ indicates the yield for formation of the $^{95\rm m}$Nb nucleus on $^{\rm nat}$Mo differs on 1.5--0.8 times from the yield for production $^{95\rm g}$Nb. This contradicts to theoretical prediction. At the same time, there is a gradual decrease in the values of $d(E_{\rm{\gamma max}})$ with an increase in the energy $E_{\rm{\gamma max}}$, which may be associated with a change in the contributions from different molybdenum isotopes to the formation of the $^{95\rm m,g}$Nb isomeric pair.

The experimental results obtained in this work and data \cite{20,21,22} significantly exceed theoretical estimates obtained with the cross-sections from the TALYS1.95 code, level density models $LD$ 1--6. Difference between theory and experiment is from 3.85 up to 5.80 times. These factors were obtained by the $\chi^2$ method with using all experimental values shown in Fig.~\ref{fig8}.   
	The result of the theoretical calculation $d(E_{\rm{\gamma max}})$, obtained using the level density model $LD$3, is closest to experimental data. 

The observed difference may be due to the underestimation of the calculation for the cross-section of the formation of the nucleus in the isomeric state. Alternatively, it may be the result of incorrect calculation of both $\sigma_{\rm m}(E)$ and $\sigma_{\rm g}(E)$. However, the theoretical prediction of the energy dependence $d(E_{\rm{\gamma max}})$ describes well the gradual decrease in the measured values with increasing $E_{\rm{\gamma max}}$ in the energy range under study.

\section{Conclusions}
\label{Concl}

In the present work, the experiment was performed using the beam from the NSC KIPT linear accelerator LUE-40 and activation and off-line $\gamma$-ray spectrometric technique. The isomeric yield ratios $d(E_{\rm{\gamma max}})$ of the $^{95\rm m,g}$Nb reaction products from photonuclear reactions on natural Mo targets were determined. The region of end-point energy of bremsstrahlung $\gamma$-quanta spectra was $E_{\rm{\gamma max}}$ = 35--95 MeV.

The data from \cite{20,21,22} are in satisfactory agreement with obtained results.

The calculation of $d(E_{\rm{\gamma max}})$ was performed using the cross-sections $\sigma(E)$ for studied reactions from the TALYS1.95 code, for six level density model $LD$ 1--6. 

Comparison of the experimental and calculated values of $d(E_{\rm{\gamma max}})$ for the reaction ${^{\rm nat}\rm{Mo}}(\gamma,x\rm np)^{95\rm m,g}\rm{Nb}$ showed a significant excess  of the experimental results over the computations. The result of the theoretical calculation $d(E_{\rm{\gamma max}})$, obtained using the level density model $LD$3, is closest to experimental data. The calculated energy dependence $d(E_{\rm{\gamma max}})$ in the energy range under study describes well the gradual decrease in the measured values with increasing energy $E_{\rm{\gamma max}}$.


\section*{Acknowlegment}
The authors would like to thank the staff of the linear electron accelerator LUE-40 NSC KIPT, Kharkiv, Ukraine, for their cooperation in the realization of the experiment.

\section*{Declaration of competing interest}
The authors declare that they have no known competing financial interests or personal relationships that could have appeared to influence the work reported in this paper.

\end{document}